\documentclass[aps,prd,twocolumn]{revtex4}
\usepackage{graphicx}
\usepackage{amsmath}
\usepackage{subfigure}
\usepackage{slashed}

\begin{document}
%=====================================================================================
%=====================================================================================
\title{Unraveling $K(1690)$ as a pseudoscalar $ud\bar{d}\bar{s}$ tetraquark state}
%=====================================================================================
%=====================================================================================
%

\author{Jin-Peng Zhang$^1$}
%\email{Zhangjp58@mail2.sysu.edu.cn}
\author{Xu-Liang Chen$^1$}
%\email{chenxliang@mail2.sysu.edu.cn}
\author{Zi-Xi Ou-Yang$^2$}
%\email{ouyangzixi23@mails.ucas.ac.cn}
\author{Xiang Yu$^1$}
%\email{yuxiang5@mail2.sysu.edu.cn}
\author{Wei Chen$^{1,3}$}
\email{chenwei29@mail.sysu.edu.cn}
\author{Jia-Jun Wu$^{2,3}$}

\affiliation{$^1$School of Physics, Sun Yat-Sen University, Guangzhou 510275, China\\
	$^2$School of Physical Sciences, University of Chinese Academy of Sciences, Beijing 100049, China\\
	$^3$Southern Center for Nuclear-Science Theory (SCNT), Institute of Modern Physics, 
Chinese Academy of Sciences, Huizhou 516000, Guangdong Province, China
}

\begin{abstract}
The recent observed $K (1690)$ has been identified as a supernumerary pseudoscalar resonance signal in the strange-meson spectrum predicted by quark model calculations. It is the best candidate of a strange crypto-exotic state. In this work, we systematically study the hadron masses of $ud\bar{d}\bar{s}$ tetraquark states with $J^P = 0^-$ in the method of QCD sum rules (QCDSR). For ten interpolating currents, we calculate the correlation functions up to dimension-8 nonperturbative condensates. To calculate the tri-gluon condensate, we comprehensively consider the contributions from different operators with and without covariant derivatives. The infrared (IR) safety can be guaranteed for the completely calculated tri-gluon condensate by properly addressing the IR divergences in Feynman diagrams. It is demonstrated that the tri-gluon condensate provides significant contributions to the sum-rule analyses in these light tetraquark systems. Our results support  the interpretation of $K (1690)$ resonance to be a pseudoscalar $ud\bar{d}\bar{s}$ tetraquark state.
\end{abstract}
\maketitle
\pagenumbering{arabic}
%
%
%
%=====================================================================================
%=====================================================================================
\section{Introduction}\label{sec:intro}
%=====================================================================================
%=====================================================================================
%
Since the establishment of the quark model (QM)~\cite{Gell-Mann:1964ewy,Zweig:1964jf}, the search for multiquark states outside of QM has always been a research hotspot in both experimental and theoretical aspects. To date, there have been many candidates of multiquark states, such as the hidden-charm pentaquarks, the charged $Z_c$ states, the doubly charged $T_{c\bar s0}(2900)^{++}$ state, the doubly charmed $T_{cc}^+$ state, etc~\cite{Chen:2016qju,Lebed:2016hpi,Esposito:2016noz,Guo:2017jvc,Liu:2019zoy,Brambilla:2019esw,Chen:2022asf,Meng:2022ozq,Liu:2024uxn,Wang:2025sic}. While most of these confirmed multiquark candidates contain heavy quarks, there also have been some notable advancements in the light hadron sector.

Recently, the COMPASS Collaboration observed three resonance structures in the P-wave $\rho (770) K$ decay channel with quantum numbers $J^P = 0^-$, which lie around $1.4 \,\text{GeV}$, $1.6 \,\text{GeV}$ and $1.8 \,\text{GeV}$ respectively~\cite{Pekeler:2024sox,Wallner:2023rmn,COMPASS:2025wkw}. However, there exist only two excited pseudoscalar strange mesons based on the quark model predictions in the mass region below $2.5\,\text{GeV}$~\cite{Ebert:2009ub,Godfrey:1985xj,Pang:2017dlw,Oudichhya:2023lva,Taboada-Nieto:2022igy}. The lightest and heaviest structures of COMPASS are consistent with the established $K (1460)$ and $K (1830)$ states in PDG~\cite{ParticleDataGroup:2024cfk} respectively, and roughly match the two states predicted by quark model calculations. 

The mass and decay width of the middle structure in COMPASS's observation are $1687\pm10^{+2}_{-67}\,\text{MeV}$ and $140\pm20^{+50}_{-50}\,\text{MeV}$ respectively, and thus dubbed $K(1690)$. This structure is not considered as the known $K(1630)$ in PDG, since the decay width is much smaller for the latter state. Moreover, the spin-parity of $K(1630)$ is still undetermined to date. 
Therefore, $K(1690)$ is a supernumerary signal in the strange meson spectrum, which can be considered as the  clear candidate for a crypto-exotic state with $J^P = 0^-$. 

In this work, we explore the possibility of $K (1690)$ as a compact tetraquark state by systematically calculating the mass spectra of $q q \bar{q} \bar{s}$ tetraquarks ($q=u/d$) with $J^P = 0^-$. In Refs.~\cite{Cotanch:2006wv,General:2007bk}, the masses of fully-light hadron molecules and tetraquark states with $J^{P C}=0^{--}$ were predicted to be around $1.36\,\text{GeV}$ and $2.1\,\text{GeV}$ respectively in the QCD Coulomb gauge approach. Using the Cornell potential-based phenomenology, the mass spectra of the S-wave and P-wave $sq\bar{q}\bar{q}$ tetraquark states were calculated in Ref.~\cite{Lodha:2025ffp}, giving quite different results for the $\bar{3}_c\otimes3_c$ and $6_c\otimes\bar{6}_c$ states. In Refs.~\cite{Jiao:2009ra,Huang:2016rro}, the $q q \bar{q} \bar{q}$ light tetraquark states with $J^{P C} = 0^{- -}$ have been studied in QCDSR. However, the calculations of tri-gluon condensate $\langle g^3fG^3 \rangle$ were incomplete and the IR divergence problem was not properly addressed in these studies. As discussed later, the contribution of tri-gluon condensate $\langle g^3fG^3 \rangle$ is very important for the light tetraquark QCDSR analyses. In this work, we shall conduct a detailed study of $\langle g^3fG^3 \rangle$ term by including all involved operators and addressing the IR divergence in the calculations. 

The paper is organized as follows. In Sec.~\ref{sec2}, we briefly introduce the method of QCDSR and construct the pseudoscalar interpolating currents for the $u d \bar{d} \bar{s}$ tetraquark states. We calculate the correlation functions and collect the results in Appendix~\ref{appendix: ope}. The details for handling the IR divergence of $\langle g^3fG^3 \rangle$ condensate are shown in Appendix~\ref{appendix: IR for G}. In Sec.~\ref{sec3}, we perform numerical
analyses and extract hadron masses of $u d \bar{d} \bar{s}$ tetraquark states. The last
section is a brief summary.

%
%=====================================================================================
%=====================================================================================
\section{QCD sum rules for light tetraquark}\label{sec2}
%=====================================================================================
%=====================================================================================
%

In QCDSR~\cite{Shifman:1978bx,Reinders:1984sr}, we consider the two-point correlation function
\begin{equation}
  \Pi(q^2)= \text{i} \int \text{d}^d x \,\text{e}^{\text{i} q · x} \langle 0|
  T [J (x) J^{\dagger} (0) |0 \rangle,
\end{equation}
where $J (x)$ is the interpolating current with the same quantum numbers to the interested physical state. Following Ref.~\cite{Jiao:2009ra}, we construct the pseudoscalar tetraquark currents with $J^P {= 0^-} $ and quark component $u d \bar{d} \bar{s}$
%%%%%%%%%%%%%%%%%%%%%%%%%%%%%%%%%%%%%%%%%%%%%%%%%%%%%%%%%%%%%%%%%%%%%%%%%%%%%%

\begin{equation}
    \begin{aligned}
    6_F \otimes \bar{6}_F (S) 
    & \left\{\begin{array}{l}
     S_6 = u_a^T C d_b [\bar{d}_a \gamma_5 C \bar{s}^T_b +
%     \bar{d}_b \gamma_5 C \bar{s}^T_a
     (a \leftrightarrow b)
     ]\\
     V_6 = u_a^T C \gamma_5 d_b [\bar{d}_a C \bar{s}^T_b +
%     \bar{d}_b C \bar{s}^T_a
      (a \leftrightarrow b)
     ]
     \\
     T_3 = u_a^T C \sigma_{\mu \nu} d_b [\bar{d}_a \sigma_{\mu \nu}
     \gamma_5 C \bar{s}^T_b - 
      (a \leftrightarrow b)
%    \bar{d}_b \sigma_{\mu \nu} \gamma_5 C
%     \bar{s}^T_a
      ]
   \end{array}\right. \\
   \bar{3}_F \otimes 3_F (A)
    &  \left\{\begin{array}{l}
     S_3 = u_a^T C d_b [\bar{d}_a \gamma_5 C \bar{s}^T_b -
%     \bar{d}_b \gamma_5 C \bar{s}^T_a
     (a \leftrightarrow b)
     ]
     \\
     V_3 = u_a^T C \gamma_5 d_b [\bar{d}_a C \bar{s}^T_b -
%     \bar{d}_b C \bar{s}^T_a
     (a \leftrightarrow b)
     ]
     \\
     T_6 = u_a^T C \sigma_{\mu \nu} d_b [\bar{d}_a \sigma_{\mu \nu}
     \gamma_5 C \bar{s}^T_b + 
     (a \leftrightarrow b)
%     \bar{d}_b \sigma_{\mu \nu} \gamma_5 C
 %    \bar{s}^T_a
     ]
   \end{array}\right. \\
  \bar{3}_F \otimes \bar{6}_F (M)
    &  \left\{\begin{array}{l}
     A_6 = u_a^T C \gamma_{\mu} d_b [\bar{d}_a \gamma_{\mu} \gamma_5 C
     \bar{s}^T_b +
     (a \leftrightarrow b)
%     \bar{d}_b \gamma_{\mu} \gamma_5 C
%     \bar{s}^T_a
     ]
     \\
     P_3 = u_a^T C \gamma_{\mu} \gamma_5 d_b [\bar{d}_a \gamma_{\mu} C
     \bar{s}^T_b -
     (a \leftrightarrow b)
%     \bar{d}_b \gamma_{\mu} C \bar{s}^T_a
     ]
   \end{array}\right. \\
  6_F \otimes 3_F (M) 
     & \left  \{   \begin{array}{l}
     P_6 = u_a^T C \gamma_{\mu} \gamma_5 d_b [\bar{d}_a \gamma_{\mu} C
     \bar{s}^T_b + 
     (a \leftrightarrow b)
     %\bar{d}_b \gamma_{\mu} C \bar{s}^T_a
     ]\\
     A_3 = u_a^T C \gamma_{\mu} d_b [\bar{d}_a \gamma_{\mu} \gamma_5 C
     \bar{s}^T_b - 
     (a \leftrightarrow b)
%     \bar{d}_b \gamma_{\mu} \gamma_5 C        %     \bar{s}^T_a
     ]   
   \end{array} \right.  ,
   \end{aligned}   
   \label{eq:currents} 
\end{equation}
in which $a, b$ are color indices, $C = \text{i} \gamma_2 \gamma_0$ is the charge conjugation operator. The flavor structures of the diquark and antidiquark operators are $6_F \otimes \bar{6}_F (S)$ for the currents $S_6, V_6, T_3$, $\bar 3_F \otimes 3_F (A)$ for the currents $S_3, V_3, T_6$, $\bar 3_F \otimes \bar 6_F (M)$ for the currents $A_6, P_3$, and $6_F \otimes 3_F (M)$ for the currents $P_6, A_3$. The subscripts 6 and 3 of the currents denote the color structures $6\otimes \bar 6$ and $\bar 3\otimes 3$, respectively. According to the flavor structures, the isospin for currents $S_6, V_6, T_3, P_6, A_3$ can be $I=3/2$ or $1/2$, while the isospin for the currents $S_3, V_3, T_6, A_6, P_3$ is $I=1/2$. 

At the hadron level, the correlation function can be described by the
dispersion relation~\cite{Chen:2024ppj}
\begin{equation}
    \begin{aligned}
 \Pi (q^2) = \frac{(q^2)^n}{\pi} \int_0^{\infty} \frac{\text{Im} \Pi
   (s)}{s^n (s - q^2)} \text{d} s + \sum_{k = 0}^{n - 1} a_n (q^2)^k,
   \end{aligned} \label{dispersionrelation}
\end{equation}
where the $a_n$ is the subtraction constant. The spectral function is usually defined as the imaginary part of correlation function
\begin{equation}
    \begin{aligned}
 \rho (s) = \frac{1}{\pi}\text{Im} \Pi (s) = f^2 \delta (s - m_X^2) + \cdots , \label{rho}
    \end{aligned}
\end{equation}
where the ``narrow resonance'' assumption is adopted in the last step, and ``$\cdots$'' contains contributions from the higher states and continuum. $f$ and $m_X$ are the coupling constant and mass of the lowest-lying hadron
state, respectively.

At the quark-gluon level, the correlation function can be computed via the method of operator product expansion
(OPE). In our calculation, we use the $d$-dimensional full quark propagator taking into account various quark and gluon condensates~\cite{Zhong:2014jla,Yang:1993bp,Grozin:1994hd}
%
%%%%%%%%%%%%%%%%%%%%%%%%%%%%%%%%%%%%%%%%%%%%%%%%%%%%%%%%%%%%%%%%%%%%%%%%%%%%%%
\begin{widetext}
    \begin{equation}
    \begin{aligned}
  S^{i j} (x) & =  \frac{\text{i} \Gamma \left( \frac{d}{2} \right)
  \slashed{x}}{2 \pi^{d / 2} (- x^2)^{d / 2}} \delta^{i j} + \frac{m \Gamma \left(
  \frac{d}{2} - 1 \right)}{4 \pi^{d / 2} (- x^2)^{d / 2 - 1}} \delta^{i j} -
  \frac{\delta^{i j}}{12} \langle \bar{\psi} \psi \rangle +
  \frac{\text{i} m \delta^{i j}}{12 d} \langle \bar{\psi} \psi \rangle
  \slashed{x} - \frac{\delta^{i j}}{48 d} \langle g \bar{\psi} \sigma G \psi
  \rangle x^2\\
  & -  \frac{\text{i} \delta^{i j} x^2 \slashed{x}}{2^4 3^4 (d + 2)} g^2 \langle
  \bar{\psi} \psi \rangle^2 + \frac{\text{i} \delta^{i j} m x^2
  \slashed{x}}{2^4 3 d (d + 2)} \langle g \bar{\psi} \sigma G \psi \rangle -
  \frac{\delta^{i j} x^4 \langle \bar{\psi} \psi \rangle \langle g^2 G^2
  \rangle}{2^6 3^2 d (d + 2)} - \frac{\delta^{i j} x^4}{2^4 3^4 d (d + 2)} g^2
  m \langle \bar{\psi} \psi \rangle^2\\
  & -  \frac{\text{i} \delta^{i j} \Gamma \left( \frac{d}{2} - 1 \right)
  \slashed{x} \langle g^3 f G^3 \rangle}{2^8 3^3 d (d + 2) \pi^{d / 2} (- x^2)^{d
  / 2 - 3}}
% - \frac{\text{i} \delta^{i j} \Gamma \left( %\frac{d}{2} - 1
%  \right) \slashed{x} \langle g^3 f G^3 \rangle}%{2^6 3^3 d (d + 2) \pi^{d / 2} (-
%  x^2)^{d / 2 - 3}}
  + \frac{\Gamma \left( \frac{d}{2} - 1 \right) \gamma^{\mu}
  \slashed{x} \gamma^{\nu}}{16 \pi^{d / 2} (- x^2)^{d / 2 - 1}} g G_{\mu \nu}^a
  T^a_{i j}\\
  & +  \left[ \frac{\Gamma \left( \frac{d}{2} - 2 \right)
  ({\gamma^{\mu \rho \nu} + \gamma^{\rho \mu \nu} - 4
  g^{\mu \rho} \gamma^{\nu}})}{96 \pi^{d / 2} (- x^2)^{d / 2 - 2}} +
  \frac{\Gamma \left( \frac{d}{2} - 1 \right) \left( x^{\mu} \gamma^{\rho}
  \slashed{x} \gamma^{\nu} + x^{\rho} \gamma^{\mu} \slashed{x} \gamma^{\nu}
  \right)}{48 \pi^{d / 2} (- x^2)^{d / 2 - 1}} \right] g G_{\mu \nu ; \rho}^a
  T^a_{i j}\\
  & +  \left( \frac{- \Gamma \left( \frac{d}{2} - 2 \right) \left(
  {2 g^{\{ \mu \rho } x^{ \sigma \}}}
  + g^{\{ \mu \rho } \gamma^{ \sigma \}} \slashed{x}
  \right)}{2^8 \times 3 \pi^{d / 2} (- x^2)^{d / 2 - 2}} + \frac{\Gamma \left(\frac{d}{2} - 1 \right) x^{\{ \mu} x^{\rho} \gamma^{\sigma \}} \slashed{x}}{192 \pi^{d / 2} (- x^2)^{d / 2 - 1}} \right)
  \gamma_{\nu} g G_{\mu \nu ; \rho \sigma}^a T^a_{i j}\\
  & +  g^2 G^a_{\mu \nu} G^b_{\rho \sigma} (T^a T^b)_{i j} \bigg[ \frac{-
  \text{i} \Gamma \left( \frac{d}{2} - 1 \right) x^{\nu} x^{\sigma}
  \gamma^{\mu} \slashed{x} \gamma^{\rho}}{96 \pi^{d / 2} (- x^2)^{d / 2 - 1}} +
  {\frac{- \text{i} \Gamma \left( \frac{d}{2} - 2 \right)
  g^{\nu \sigma} \gamma^{\mu} \slashed{x} \gamma^{\rho}}{192 \pi^{d / 2} (-
  x^2)^{d / 2 - 2}}}  \\
  &+  \frac{ \text{i} \Gamma \left( \frac{d}{2} - 2 \right)}{2^8 \times 3
  \pi^{d / 2} (- x^2)^{d / 2 - 2}} ( - 6 g^{\nu \sigma} \slashed{x}
  \gamma^{\mu \rho} - 4 x^{\sigma} \gamma^{\mu \nu \rho} + 6 x^{\mu}
  \gamma^{\nu \rho \sigma} - 4 x^{\nu} \gamma^{\mu \sigma \rho}
  {+ 3 \gamma^{\mu} \slashed{x} \gamma^{\nu \rho \sigma}})\bigg] ,
    \end{aligned}
    \label{eq:chuanboz}
\end{equation}
\end{widetext}
%%%%%%%%%%%%%%%%%%%%%%%%%%%%%%%%%%%%%%%%%%%%%%%%%%%%%%%%%%%%%%%%%%%%%%%%%%%%%%
%
where $m$ is the light quark mass, $i, j=1, 2, 3$ are color indices, $T^a (a=1,...,8)$ is the Gell-Mann matrix. We also use the following abbreviations
\begin{widetext}
\begin{equation}
    \begin{aligned}
  \gamma^{\mu \nu \rho} & :=  \gamma^{\mu} \gamma^{\nu} \gamma^{\rho}, \quad
  g^{\{ \mu \rho } x^{ \sigma \}}  :=  g^{\mu \rho}
  x^{\sigma} + g^{\mu \sigma} x^{\rho} + g^{\rho \sigma} x^{\mu}, \quad
  g^{\{ \mu \rho } \gamma^{ \sigma \}}  :=  g^{\mu \rho}
  \gamma^{\sigma} + g^{\mu \sigma} \gamma^{\rho} + g^{\rho \sigma}
  \gamma^{\mu},  \\
  x^{\{ \mu } & x^{\rho} \gamma^{ \sigma \}}  :=  x^{\mu}
  x^{\rho} \gamma^{\sigma} + x^{\mu} x^{\sigma} \gamma^{\rho} + x^{\rho}
  x^{\sigma} \gamma^{\mu}, \quad
  G_{\mu \nu ; \rho}^a  := \tilde{D}_\rho^{ab} G^b_{\mu \nu}, \quad
  G_{\mu \nu ; \rho \sigma}^a  := \tilde{D}_\sigma^{a b} \tilde{D}_\rho^{b c}  G^{c}_{\mu \nu},
    \end{aligned}
\end{equation}
\end{widetext}
in which $G_{\mu \nu}$ is the gluon filed strength, $\tilde{D}_\mu^{a b}=\delta^{a b}\partial_\mu-g f^{a b c} A^c_\mu$ is the covariant derivative operator in the adjoint representation and $A^c_\mu$ is the external gauge field. The origin of the full quark propagator in Eq.~\eqref{eq:chuanboz} includes both the background field and quantum field. In our calculation, we keep track of terms in the fifth order of the background quark field expansion $\frac{x^{\mu}x^{\nu} x^{\rho} x^{\sigma}}{4!} \langle \psi^i D_{\mu} D_{\nu} D_{\rho}
D_{\sigma} \bar{\psi}^j \rangle$. 
For the part of quantum fields, we include terms up to dimension-6 to fully compute the result of $\langle g^3 f G^3 \rangle$.

%%%%%%%%%%%%%%%%%%%%%%%%%%%%%%%%%%%%%%%%%%%%%%%%%%%%%%%%%%%%%%%%%%%%%%%%%%%%%%
%
In QCDSR studies for multiquark systems, the contribution of tri-gluon condensate $\langle
g^3 f G^3 \rangle$ is usually neglected due to the computational complexity and infrared divergence (IR) in fully calculations. 
%An exception is Ref.\cite{Li:2024rrs}, but they did not %summarize the cancellation of the IR divergence of $%\langle g^3 f G^3\rangle$ nor provide a detailed %analysis of its contribution. 
However, such contributions may be significant in fully  calculations at leading order (LO) by considering $\langle (D G) (D G)
\rangle$ and $\langle G (D D G) \rangle$ terms as the following~\cite{Zhong:2014jla,Chen:2013zia}
\begin{equation}
    \begin{aligned}
   &\langle  G^a_{\mu \nu ;\rho}  G^b_{\alpha \beta ;  \sigma} \rangle  = 
   \delta^{a b} 2Z
  g_{\rho \sigma} (g_{\mu \alpha} g_{\beta \nu} - g_{\alpha \nu} g_{\beta
  \mu}) \\
  & + \delta^{a b} Z  [g_{\nu \sigma} g_{\mu \alpha} g_{\beta \rho} + g_{\mu \sigma}
   g_{\nu \beta} g_{\alpha \rho} - (\mu \leftrightarrow \nu)]  
   \\
  & + \delta^{a b} Y [(g_{\nu \rho} g_{\mu \alpha} g_{\beta \sigma} + g_{\mu
  \rho} g_{\nu \beta} g_{\alpha \sigma}) - (\mu \leftrightarrow \nu)],
  \label{eq:njl}\\
  Z & =  - \frac{D (D - 2) g^2 \langle \bar{\psi} \psi \rangle^2 + 9 (D
  - 1) \langle g f G^3 \rangle}{72 (D - 2) (D - 1) D (D + 2)},\\
  Y & =  \frac{- D (D - 2) g^2 \langle \bar{\psi} \psi \rangle^2 + 27
  \langle g f G^3 \rangle}{72 (D - 2) (D - 1) D (D + 2)},
    \end{aligned}
\end{equation}
\begin{equation}
    \begin{aligned}
\langle G^a_{\mu \nu ; \rho}  G^b_{\alpha \beta ; \sigma} \rangle =  -
  \langle  G^a_{\mu \nu ; \rho \sigma} G^b_{\alpha \beta} \rangle = - \langle
  G^a_{\mu \nu} G^b_{\alpha \beta ; \sigma \rho} \rangle,
  \end{aligned} \label{eq:njl11}
\end{equation}
in which we adopt the $D$-dimensional condensates to  distinguish the dimension $d$ of the full quark propagator in Eq.~\eqref{eq:chuanboz}. $D$ and $d$ are identical, but such symbols are useful for addressing IR divergences, as shown in Appendix~\ref{appendix: IR for G}. In Ref.~\cite{Li:2024rrs}, the authors also address the IR divergence of $\langle g^3 f G^3\rangle$, they just did not summarize the corresponding cancellation tricks. One should note that the IR safety for the tri-gluon condensate $\langle g^3 f G^3 \rangle$ can not be guaranteed for calculating the $\langle (D G) (D G)
\rangle$ terms by using propagators up to dimension-6, as adopted in Ref.~\cite{Zhong:2014jla}.
\begin{figure*}[t]
    \centering
        \includegraphics[width=0.78\textwidth]{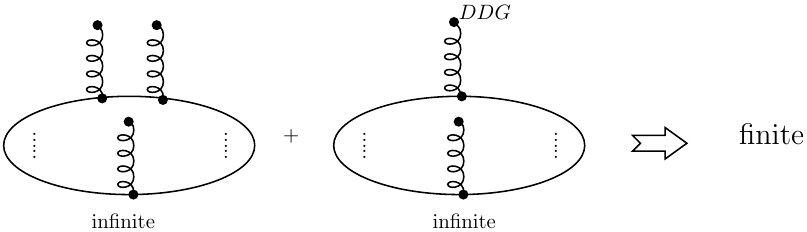}
    \caption{Cancellation of $\langle (GG)G\rangle$ and $\langle  G(DDG)\rangle$ condensates formed by two different quark propagators.}
    \label{fig:IR1}
\end{figure*}
\begin{figure*}
	\centering
	\subfigure[]{
	\includegraphics[width=0.39\textwidth]{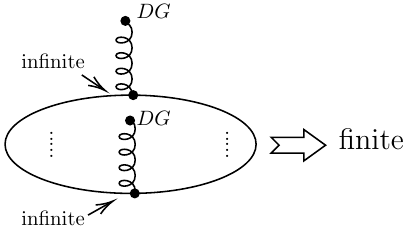}}
	\qquad
	\subfigure[]{
	\includegraphics[width=0.39\textwidth]{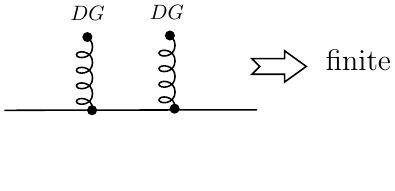}}
	\caption{IR safety for $\langle (DG)(DG)\rangle$ condensate: (a) cancellation of $\langle (DG)(DG) \rangle$ by any two quark propagators; (b) the propagator containing $\langle (DG)(DG) \rangle$ has no IR divergence.}
	\label{fig:IR2}
\end{figure*}
%\begin{figure*}[t]
%    \centering
 %   \subfloat[]{
  %      \includegraphics[width=0.39\textwidth]{IR2.1.pdf}
  %      }\qquad
  %  \subfloat[]{
  %      \includegraphics[width=0.39\textwidth]{IR2.2.pdf}
  %      }\qquad
  %  \caption{IR safety for $\langle (DG)(DG)\rangle$ condensate: (a) cancellation of $\langle (DG)(DG) \rangle$ by any two quark propagators; (b) the propagator containing $\langle (DG)(DG) \rangle$ has no IR divergence.}
  %  \label{fig:IR2}
%\end{figure*}

\begin{figure*}[htbp]
    \centering
        \includegraphics[width=0.78\textwidth]{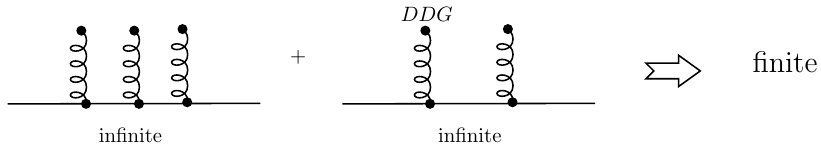}
    \caption{Cancellation of $\langle GGG \rangle$ and $\langle G(DDG)\rangle$ condensates formed by the single quark propagator.}
    \label{fig:IR3}
\end{figure*}

For the full quark propagator in Eq.~\eqref{eq:chuanboz}, the IR divergences appear for the terms involving $\Gamma (d/2 - 2)$ when $d=4$. Fortunately, the operator mixing is
unnecessary for $\langle g^3 f G^3 \rangle$ to address the IR divergence if we neglect the $g^2 \langle\bar{\psi} \psi \rangle^2$ parts in the expressions of $\langle (D G) (D G) \rangle$ and
$\langle G (D D G) \rangle$~\cite{Broadhurst:1984rr}, which are actually of high order in $\alpha_s$ and thus numerically negligible. Therefore, its
contribution can be neglected. We summarize the handling of the IR divergence
for $\langle g^3 f G^3 \rangle$ as follows:
\begin{itemize}
  \item As shown in Fig.~\ref{fig:IR1}, both the $\langle G (D D G) \rangle$ and $\langle G (G G) \rangle$ condensates formed by two different quark propagators are divergent. However, the divergences from these two diagrams cancel each other out exactly.
  \item Although the propagator including $D G$ is divergent, the $\langle (D
  G) (D G) \rangle$ condensate in any diagram has no IR divergence, as shown in Fig.~\ref{fig:IR2}.
  \item As shown in Fig.~\ref{fig:IR3}, both the $\langle G (D D G) \rangle$ and $\langle G G G \rangle$ condensates formed by the single quark propagator are divergent. However, the divergences from these two diagrams cancel each other out exactly.
\end{itemize}
The corresponding propagators are as follows:
\begin{equation}
    \begin{aligned}
   S^{ij}_{\langle (DG) (DG) \rangle} (x)&= \frac{- \text{i}
   \delta^{i j} \langle g^3 f G^3 \rangle \Gamma \left( \frac{d}{2} - 1
   \right) \slashed{x}}{1728 d (d + 2) \pi^{d / 2} (- x^2)^{d / 2 - 3}} ,
   \end{aligned}
\end{equation}
\begin{equation}
    \begin{aligned}
   S^{ij}_{\langle GGG \rangle} (x) &= \frac{- \text{i} \delta^{i j}
   \langle g^3 f G^3 \rangle \Gamma \left( \frac{d}{2} - 2 \right)
   \slashed{x}}{9216 d \pi^{d / 2} (- x^2)^{d / 2 - 3}} ,
   \end{aligned}
\end{equation}
\begin{equation}
    \begin{aligned}
   S^{ij}_{\langle G (DDG) \rangle} (x) &= \frac{\text{i} \delta^{i j} (d - 2)
    \langle g^3 f G^3 \rangle \Gamma \left( \frac{d}{2} - 2
   \right) \slashed{x}}{3072 d (d + 2) \pi^{d / 2} (- x^2)^{d / 2 - 3}} ,
   \end{aligned}
\end{equation}
in which the last two ones are divergent while the sum of them are finite. The detailed discussions can be found in  Appendix~\ref{appendix: IR for G}. 
After considering all the above points, we can completely calculate the contribution of tri-gluon condensate $\langle g^3 f G^3 \rangle$ at the leading order (LO) of $\alpha_s$. 

\begin{figure*}
	\centering
	\subfigure[]{
		\label{fig:Feynman4a}
		\includegraphics[width=0.23\textwidth]{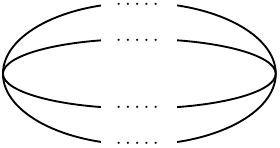}}
	\subfigure[]{
		\label{fig:Feynman4b}
		\includegraphics[width=0.23\textwidth]{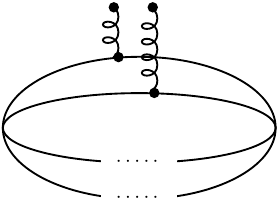}}
	\subfigure[]{
		\label{fig:Feynman4c}
		\includegraphics[width=0.23\textwidth]{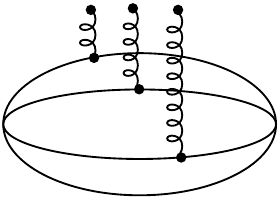}}
		 \\
	\subfigure[]{
		\label{fig:Feynman4d}
		\includegraphics[width=0.23\textwidth]{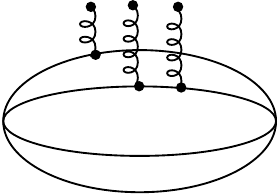}}
	\subfigure[]{
		\label{fig:Feynman4e}
		\includegraphics[width=0.23\textwidth]{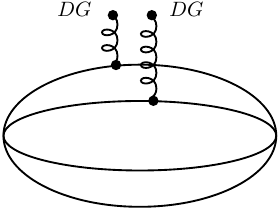}}
	\subfigure[]{
		\label{fig:Feynman4f}
		\includegraphics[width=0.23\textwidth]{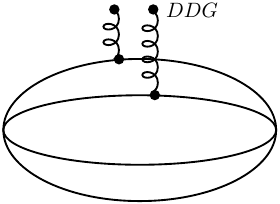}}
		 \\
	\subfigure[]{
		\label{fig:Feynman4g}
		\includegraphics[width=0.23\textwidth]{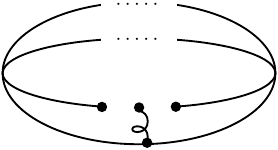}}
	\subfigure[]{
		\label{fig:Feynman4h}
		\includegraphics[width=0.23\textwidth]{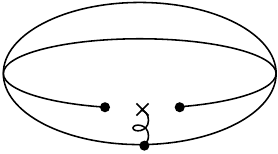}}
	\subfigure[]{
		\label{fig:Feynman4i}
		\includegraphics[width=0.23\textwidth]{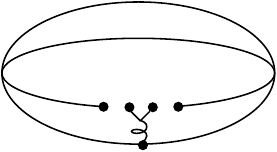}}
	\caption{The Feynman diagrams involved in our calculations for the $ud\bar{d}\bar{s}$ tetraquark systems. A quark line with dots contains terms with the color factor $\delta^{i j}$ in $S^{i j} (x)$.}
	\label{fig:Feynman}
\end{figure*}

We calculate the correlation functions up to dimension-8 by considering the Feynman diagrams depicted in Fig.~\ref{fig:Feynman}, including the complete contribution of $\langle
g^3 f G^3 \rangle$ at LO. For simplicity, we use a quark line with dots to
represent the terms with the color factor $\delta_{i j}$ in the quark propagator $S_{i j} (x)$ in Eq.~\eqref{eq:chuanboz}, as they can be calculated collectively. Gluon lines labeled $D G$ and $D D G$ correspond to terms involving $G_{\mu \nu ;
\rho}$ and $G_{\mu \nu ; \rho \sigma}$ in Eq.~\eqref{eq:chuanboz}, respectively. The IR divergences in the diagrams Fig.~\ref{fig:Feynman4d}-Fig.~\ref{fig:Feynman4f} contributing to $\langle g^3 f G^3 \rangle$ can be systematically addressed by applying the cancellation tricks presented in Figs.~\ref{fig:IR1}-\ref{fig:IR3}. The diagram Fig.~\ref{fig:Feynman4h}
incorporates contributions from two types of condensates: $\langle \bar{q} G_{\mu
\nu} D_{\rho} q \rangle$, $\langle \bar{q} \overleftarrow{D}_{\rho}
G_{\mu \nu} q \rangle$. The diagram Fig.~\ref{fig:Feynman4i} contains contribution from condensate $\langle \bar{q} G_{\mu \nu ;\rho} q \rangle$.
The results of the correlation functions are listed in Appendix~\ref{appendix: ope}.

The correlation function obtained at the quark-gluon level is expected to be
equivalent to the one at the hadron level due to the principle of quark-hadron duality. In QCDSR, one can pick out the lowest lying resonance by performing the Borel transform to the correlation functions at both levels, in order to suppress the contributions from high excited states in Eq.~\eqref{rho} and eliminate the unknown subtraction terms in Eq.~\eqref{dispersionrelation}. Subsequently, the QCD sum rules are obtained as
\begin{equation}
  \Pi (M_B^2, s_0) = f^2 e^{- m_X^2 / M^2_B} = \int_0^{s_0} \rho (s) e^{- s
  / M^2_B} \text{d} s,
\end{equation}
in which $s_0$ is the continuum threshold parameter, and $M_B$ is the Borel parameter.
Then the hadron mass $m_X$ of the lowest lying resonance can be extracted to be 
\begin{equation}
  m_X = \sqrt{\frac{L_1 (s_0, M^2_B)}{L_0 (s_0, M^2_B)}},
\end{equation}
where 
\begin{equation}
  L_k (s_0, M^2_B) = \int_0^{s_0} \rho (s) s^k e^{- s / M^2_B} \text{d} s.
\end{equation}

%\begin{figure*}[htbp]
%    \centering
%    \subfloat[]{
%        \includegraphics[width=0.23\textwidth]{pert.pdf} \label{fig:Feynman4a}
%        }
%    \subfloat[]{
%        \includegraphics[width=0.23\textwidth]{GG.pdf} \label{fig:Feynman4b}
%        }
%    \subfloat[]{
%        \includegraphics[width=0.23\textwidth]{GGG.pdf} \label{fig:Feynman4c}
%        }  \\
%    \subfloat[]{
%        \includegraphics[width=0.23\textwidth]{GGG2.pdf} \label{fig:Feynman4d}
%        }
%    \subfloat[]{
%        \includegraphics[width=0.23\textwidth]{DGDG.pdf} \label{fig:Feynman4e}
%        }
%    \subfloat[]{
%        \includegraphics[width=0.23\textwidth]{GDDG.pdf} \label{fig:Feynman4f}
%        } \\
%    \subfloat[]{
%        \includegraphics[width=0.23\textwidth]{qGq.pdf} \label{fig:Feynman4g}
%        }
%    \subfloat[]{
%        \includegraphics[width=0.23\textwidth]{qGDq.pdf} \label{fig:Feynman4h}
%        }
%    \subfloat[]{
%        \includegraphics[width=0.23\textwidth]{qDGq.pdf} \label{fig:Feynman4i}
%        }
%    \caption{The Feynman diagrams involved in our calculations for the $ud\bar{d}\bar{s}$ tetraquark systems. A quark line with dots contains terms with the color factor $\delta^{i j}$ in $S^{i j} (x)$.}
%    \label{fig:Feynman}
%\end{figure*}
%=====================================================================================
%=====================================================================================
\section{Numerical analysis}\label{sec3}
%=====================================================================================
%=====================================================================================
%
To perform the QCD sum rule numerical analyses, we use the following values of the quark masses, strong coupling and various QCD condensates~\cite{ParticleDataGroup:2024cfk,Jin:2002rw,Huang:2014hya,Narison:2025cys,Narison:2011rn,Narison:2011xe,DiGiacomo:1982gn,Shifman:1978by}:

\begin{align}
	&\qquad  m_u = m_d = 0  ,  \quad m_s  =  93.5 \pm 0.8 \,\text{MeV},\nonumber \\
	& \langle \bar{q} q \rangle = - (0.24 \pm 0.01)^3 \,\text{GeV}^3  , 
	\langle \bar{s} s \rangle = (0.8 \pm 0.1) \langle \bar{q} q
	\rangle, \nonumber \\
	&\qquad \quad  \langle g \bar{q} \sigma G q \rangle = (0.8 \pm 0.2)  \langle
	\bar{q} q \rangle \,\text{GeV}^2, \nonumber \\
	&\qquad \quad \langle g \bar{s} \sigma G
	s \rangle= (0.8 \pm  0.2)  \langle g \bar{q} \sigma G q \rangle, \nonumber \\
	&\qquad \langle \alpha_s G^2 \rangle  = (6.39 \pm  0.35) \times  10^{- 2}
	\,\text{GeV}^4,  \label{eq:para} \\
	&\qquad \qquad \langle g^3 f G^3 \rangle  = 1.2 \langle \alpha_s  G^2 
	\rangle  \,\text{GeV}^2 , \nonumber \\
	&\qquad    m_s (\mu)  = m_s (2 \,\text{GeV})  \left[ \frac{\alpha_s (\mu)}{\alpha_s (2
		\,\text{GeV})} \right]^{\frac{12}{33 - 2 n_f}}, \nonumber 
	\\
	& \alpha_s  = \frac{1}{b_0 t}  \left[ 1 -  \frac{b_1}{b_0^2} \frac{\log t}{t}   +    
	\frac{b_1^2  (\log^2 t - \log t - 1) + b_0 b_2}{b_0^4 t^2} \right] , \nonumber
\end{align}
where $t = \log \frac{\mu^2}{\Lambda^2}$, $b_0 = \frac{33 - 2 n_f}{12 \pi}$,
$b_1 = \frac{153 - 19 n_f}{24 \pi^2}$, $b_2 = \frac{2857 - \frac{5033}{9} n_f
+ \frac{325}{27} n_f^2}{128 \pi^3}$, $\Lambda = 332 \,\text{MeV}$ for the
flavors $n_f = 3$. The $s$-quark mass $m_s$ is taken as the
$\overline{\text{MS}}$ mass at the scale $\mu = 2 \,\text{GeV}$. Given that the
condensate values are evaluated at the renormalization scale $\mu = 1$GeV, we
accordingly adopt $m_s$ at the same scale. To date, there is no reliable method to determine the value of $\langle g^3 f G^3 \rangle$ from experimental data. In our analyses, we adopt the value from the early review papers of QCDSR~\cite{Shifman:1978bx,Reinders:1984sr,Shifman:1978by}, which was taken from estimations of instanton models~\cite{Shifman:1978by} and lattice calculation~\cite{DiGiacomo:1982gn}.
%The value of $\langle g^3 f G^3 \rangle$ is obtained from lattice calculations~\cite{DiGiacomo:1982gn} and instanton models~\cite{Shifman:1978by}. Both relate $\langle g^3 f G^3 \rangle$ to $\langle \alpha_s G^2 \rangle$. Therefore, the uncertainty of $\langle g^3 f G^3 \rangle$ originates from $\langle \alpha_s G^2 \rangle$ within this framework.

\begin{figure*}
	\centering
	\subfigure[]{
		\label{fig: contribution of different condensates}
		\includegraphics[width=0.45\textwidth]{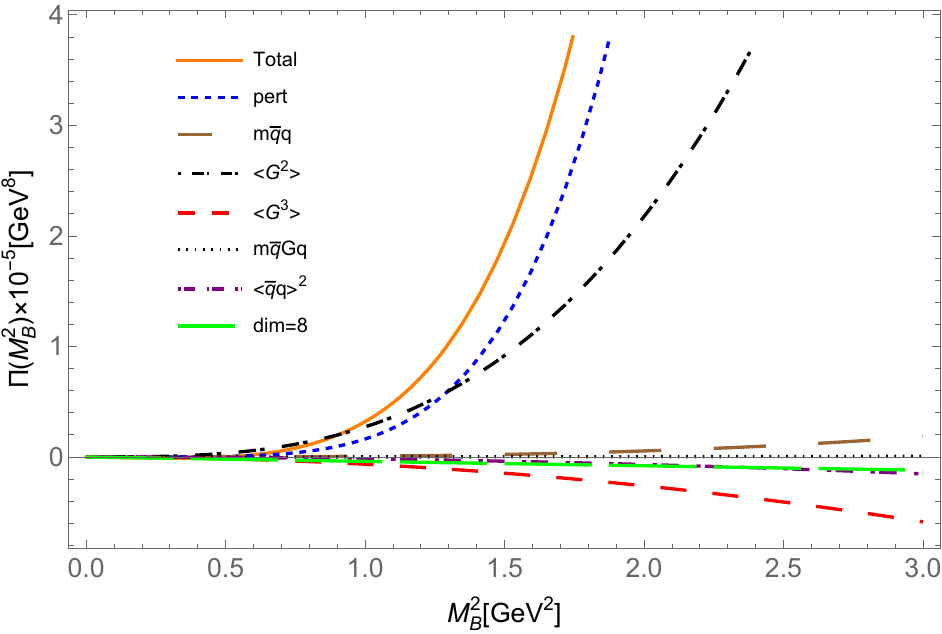}}
	\qquad
	\subfigure[]{
		\label{fig: s0}
		\includegraphics[width=0.45\textwidth]{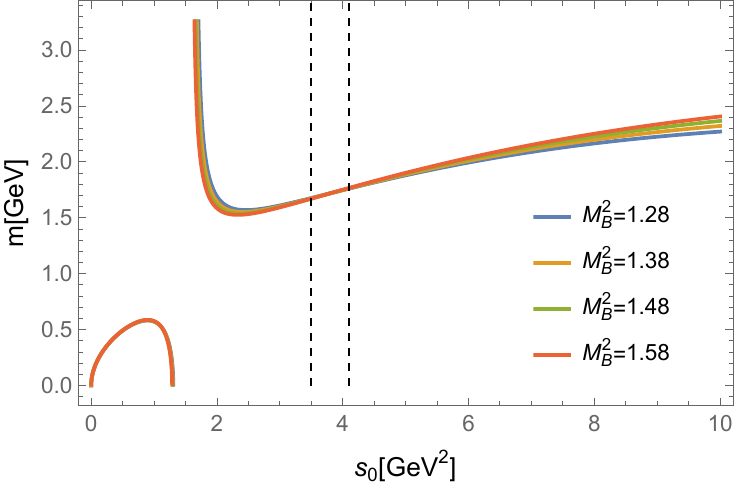}}
	\caption{The OPE convergence (a) and variation of hadron mass with $s_0$ (b) for the interpolating current $P_3$.}
	\label{fig: convergence and s0}
\end{figure*}

We first study the behaviors of correlation function to show the importance of the tri-gluon condensate $\langle g^3 f G^3 \rangle$, by taking the interpolating current $P_3$ as an example in Fig.~\ref{fig: contribution of different condensates}. The gluon condensates $\langle g^2 G^2 \rangle$ and $\langle g^3 f G^3 \rangle$ are shown to be the most important nonperturbative terms, while the quark condensates $m_s\langle \bar qq \rangle$ and $m_s\langle \bar qGq \rangle$ are much smaller due to the light quark mass. However, the four-quark condensate $\langle \bar qq \rangle^2$ also gives significant contribution to the correlation function~\cite{Shifman:1978bx,Jiao:2009ra}. Similar situations occur for other tetraquark currents in Eq.~\eqref{eq:currents}. In other words, the well-calculated tri-gluon condensate $\langle g^3 f G^3 \rangle$ may be important to the fully-light tetraquark sum rule analyses and thus cannot be neglected. This is very different from the heavy tetraquark systems, in which the quark condensates are usually much larger than gluon condensates due to the existence of heavy quarks~\cite{Nielsen:2009uh,Chen:2016qju}. The contributions of tri-gluon condensates have been proven to be negligible for the conventional heavy quarkonium~\cite{Nikolaev:1981ff,Nikolaev:1982rq} and fully-heavy baryons~\cite{Wu:2021tzo}.

\begin{figure*}
	\centering
	\subfigure[]{
		\label{fig: PC}
		\includegraphics[width=0.45\textwidth]{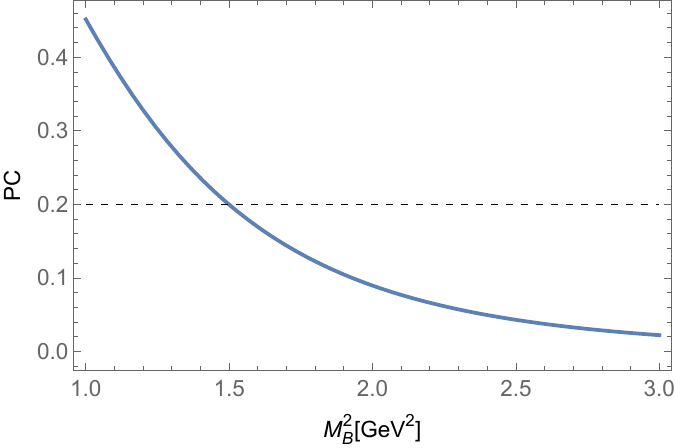}}
	\qquad
	\subfigure[]{
		\label{fig: mass}
		\includegraphics[width=0.45\textwidth]{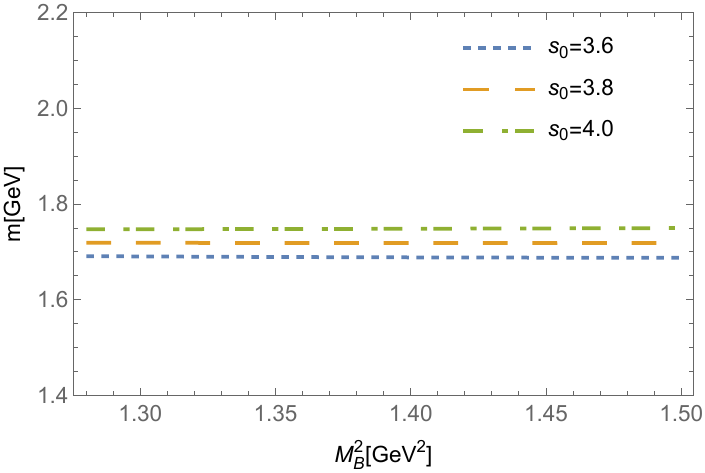}}
	\caption{The pole contribution (a) and variation of hadron mass with $M_B^2$ (b) for the interpolating current $P_3$.}
	\label{fig: PC and mass}
\end{figure*}

In QCDSR, the study of OPE convergence leads to the lower limit $M^2_{B\min}$ of the
Borel parameter. To ensure a good OPE convergence, we require that the $D = 6$ and $D = 8$ condensates give small contributions to the correlation functions
\begin{equation}
    \left| \frac{\Pi_{D = 6} (M_B^2)}{\Pi_{\text{pert}} (M_B^2)}
   \right| \le 25 \text{\%}, 
    \left| \frac{\Pi_{D = 8} (M_B^2)}{\Pi_{\text{pert}} (M_B^2)}
   \right| \le 10 \text{\%} . \label{eq: convergence of OPE}
\end{equation}
Meanwhile, the guarantee of pole contribution (PC) will provide an upper limit for $M_B^2$
\begin{equation}
    \text{PC} = \frac{L_0 (s_0, M^2_B)}{L_0 (\infty, M^2_B)} \ge 10
   \text{\%} .
\end{equation}
According to Eq.~\eqref{eq: convergence of OPE}, the behavior of OPE series in Fig.\ref{fig: contribution of different condensates} exhibits very good convergence for $M^2_B\ge 1.28\,\text{GeV}^2$. As shown in Fig.~\ref{fig: s0}, we further study the dependence of the hadron mass on the continuum threshold parameter $s_0$ with different values of $M^2_B$, from which one immediately realizes an optimal value of $s_0=3.8\pm0.2\,\text{GeV}^2$.
Within this choice, the hadron mass is almost independent of the unphysical parameter $M^2_B$. The uncertainty of the threshold value $s_0$ is estimated to be $\pm0.2\,\text{GeV}^2$, within which the stability of mass sum rules can be guaranteed.
In Fig.~\ref{fig: PC}, we show the pole contribution to set the upper limit as $M^2_{B\max} = 1.5\,\text{GeV}^2$ by requiring $\text{PC}\ge 20$\% for this current. Within these parameter working regions, the mass curves can be plotted in Fig.~\ref{fig: mass}. The hadron mass is reading as
\begin{equation}
 m_X = 1.72 \pm 0.10 \,\text{GeV} ,\label{eq: easy mass}
 \end{equation}
in which the error is from the uncertainties of $s_0$,  the strange quark mass $m_s$ and various QCD condensates. 
After performing numerical analyses for all interpolating currents in Eq.~\eqref{eq:currents}, we present the numerical results for these $ud\bar d\bar s$ tetraquarks 
in Table~\ref{table: mass for all}. To show the importance of tri-gluon condensate $\langle g^3 f G^3 \rangle$, we reanalyze the mass sum rules without considering the contributions from this term for the currents $P_6$ and $P_3$. As shown in Table~\ref{table: mass for comparison}, the predicted masses are about 20\% lower than those in Table~\ref{table: mass for all}.
\begin{figure*}[t]
	\centering
	\includegraphics[width=0.4\textwidth]{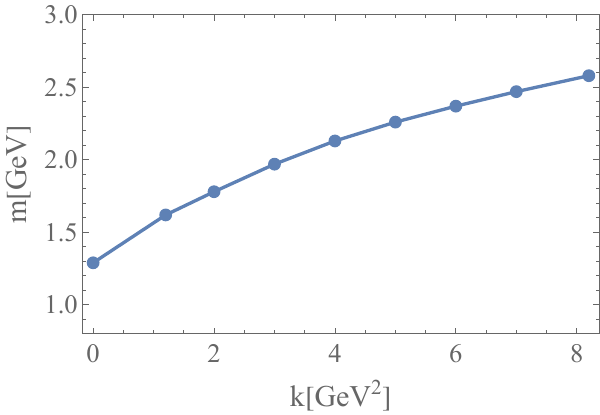}
	\caption{The sensitivity of extracted hadron mass to coefficient $k=\langle g^3 f G^3 \rangle/\langle \alpha_s G^2\rangle$ for the interpolating current $P_6$}
	\label{fig:sensitivity}
\end{figure*}
\begin{table*}[htbp] 
	\centering
	\renewcommand\arraystretch{2}
	\setlength{\tabcolsep}{6mm}
	{
		\begin{tabular}{ccccc}
			\hline\hline
			Currents   & $s_{0}(\pm0.2\,\text{GeV}^2)$ & $M_B^2~(\text{GeV}^2)$ & $m_X~(\text{GeV})$ & PC(\text{\%})\\
			\hline
			$P_6$ & $3.3$ & $1.14 \sim 1.35$ & $1.62 \pm 0.10$ & $>15$\\
			$P_3$ & $3.8$ & $1.28 \sim 1.50$ & $1.72 \pm 0.10$ & $>20$\\
			$S_6$ & $6.7$ & $1.40 \sim 1.80$ & $2.34 \pm 0.11$ & $>25$\\
			$S_3$ & / & / & / & /\\
			$A_6$ & $1.0$ & $0.54 \sim 0.62$ & $0.90 \pm 0.50$ & $>10$\\
			$A_3$ & $2.8$ & $0.95 \sim 1.20$ & $1.48 \pm 0.20$ & $>10$\\
			$V_6$ & $8.8$ & $1.34 \sim 1.94$ & $2.60 \pm 0.20$ & $>40$\\
			$V_3$ & $3.3$ & $1.42 \sim 1.62$ & $1.60 \pm 0.16$ & $>10$\\
			$T_6$ & $2.4$ & $0.75 \sim 1.10$ & $1.36 \pm 0.05$ & $>15$\\
			$T_3$ & $4.7$ & $0.90 \sim 1.35$ & $1.92 \pm 0.04$ & $>25$\\
			\hline\hline
		\end{tabular}
	}
	\caption{Numerical results for the $ud\bar{d}\bar{s}$ tetraquark states.}
	\label{table: mass for all}
\end{table*}
\begin{table*}[htbp] 
	\centering
	\renewcommand\arraystretch{2}
	\setlength{\tabcolsep}{6mm}
	{
		\begin{tabular}{cccccc}
			\hline\hline
			Currents & $s_0 (\pm0.2\,\text{GeV}^2)$ & $M_{B}^2
			(\text{GeV}^2)$ & $m_X (\text{GeV})$ & $\text{PC} \left( \text{\%} \right)$
			& $\text{R} \left( \text{\%} \right)$\\
			\hline
			$P_6$ & 2.1 & $0.88 \sim 0.97$ & $1.29 \pm 0.12$ & $>15$ & $25.6$\\
			$P_3$ & $2.7$ & $1.25 \sim 1.33$ & $1.45 \pm 0.11$ & $>15$ & $18.6$\\
			\hline\hline
		\end{tabular}
	}
	\caption{Numerical results for $P_6$ and $P_3$ currents without contribution of $\langle g^3 f G^3\rangle$. The value $R$ in the last column represents the mass correction comparing to the results in Table~\ref{table: mass for all}. }
	\label{table: mass for comparison}
\end{table*}

There have been some other discussions regarding the value of $\langle g^3 f G^3 \rangle$. For example, QCDSR provided a quite different estimation from the studies of heavy quarkonium systems~\cite{Narison:2011rn,Narison:2011xe}
\begin{equation}
	\langle g^3 f G^3 \rangle =( 8.2 \pm 1.0 ) \,\text{GeV}^2 \times \langle \alpha_s G^2\rangle ,
\end{equation}
which is much larger than that in Eq.~\eqref{eq:para}. Since the tri-gluon condensate will contribute significantly to our sum rule analyses, we further investigate the dependence of extracted hadron mass to the coefficient $k=\langle g^3 f G^3 \rangle/\langle \alpha_s G^2\rangle$ in Fig.~\ref{fig:sensitivity}, by using the interpolating current $P_6$ as an example. It is clear that the hadron mass is very sensitive to the change of the tri-gluon condensate.

In these results, the hadron masses extracted from the currents $P_6$, $P_3$ and
$V_3$ are consistent with the mass of $K (1690)$ observed by the COMPASS Collaboration~\cite{Pekeler:2024sox,Wallner:2023rmn,COMPASS:2025wkw}, suggesting that $K (1690)$ could be a $ud\bar d\bar s$ tetraquark state. 
For the currents $S_6$ and $V_6$, the dominant nonperturbative contributions are negative, leading to poor positivity behavior of the spectral functions, and thus the mass predictions for these two currents are actually unreliable. The analysis for current $S_3$ fails to produce stable mass sum rules. Moreover, the hadron masses extracted from the currents $A_3$ and $T_3$ are close to the masses of $K (1460)$ and $K(1830)$ respectively, implying possible tetraquark components in these two strange mesons. More properties of the $ud\bar d\bar s$ tetraquarks should be investigated to further study the inner structures of these resonance states.

%=====================================================================================
%=====================================================================================
\section{Summary}\label{sec5}
%=====================================================================================
To study the inner structure of the recent observed strange meson $K(1690)$, we investigate the mass spectrum of the light tetraquark $u d \bar{d} \bar{s}$ states with quantum numbers $J^P =
0^-$ in the method of QCDSR. We calculate the two-point correlation functions up to dimension-8 nonperturbative condensates at the leading order of $\alpha_s$. We utilize the quark propagators incorporating $D G$, $D D G$ and
$\langle g^3 f G^3 \rangle$ terms to comprehensively account for the contributions of tri-gluon condensate $\langle g^3 f G^3
\rangle$. After properly addressing the IR divergences in OPE, our calculations demonstrate 
that the tri-gluon condensate $\langle g^3 f G^3 \rangle$ provides significant contribution to the mass sum rule analyses for light
tetraquark systems, potentially surpassing the contributions of four-quark condensate $\langle
\bar{q} q \rangle^2$. In the case of $P_6$ and $P_3$, the inclusion of $\langle g^3 f G^3 \rangle$ can increase about 20\% of mass predictions.

Using the interpolating currents $(P_6, P_3, V_3)$, we predict the hadron mass of pseudoscalar $u d \bar{d} \bar{s}$ tetraquark state to be $1.6 - 1.7 \,\text{GeV}$. This result is consistent with the mass of $K (1690)$ observed by COMPASS, supporting the compact tetraquark interpretation of the structure. However, one should note that other possibility can not be excluded, such as a strange hybrid meson, or a structure of kinematic effect~\cite{Guo:2019twa,COMPASS:2020yhb}. The decay properties of $u d \bar{d} \bar{s}$ tetraquark state should be studied to further investigate the inner structure of $K (1690)$, such as the decay processes $u d \bar{d} \bar{s}\to a_0 K, f_0 K, K^{*}_{0}\pi, K^{*}\rho, K\rho, K^*\pi$, etc. Especially in the isospin quartet ($I=3/2$), the doubly-charged $u u \bar{d} \bar{s}$ tetraquark can be considered as a characteristic signal in the future.

%
%=====================================================================================
%=====================================================================================
%=====================================================================================
\section*{Acknowledgments}
%=====================================================================================
%=====================================================================================
%=====================================================================================
%

Jin-Peng Zhang thanks Ding-Kun Lian for useful discussions. This work is supported by the National Natural Science Foundation of China under Grant No. 12175318.

\appendix
\section{IR safety for $\langle g^3 f G^3 \rangle$}\label{appendix: IR for G}
In this calculation, the IR divergences can appear in the Feynman diagrams Figs.~\ref{fig:Feynman4d}-\ref{fig:Feynman4f}, in which the involved condensates $\langle (GG)G\rangle$, $\langle  G(DDG)\rangle$ and $\langle  (DG) (DG)\rangle$ come from any two propagators. 
 This can be expressed as
\begin{equation}
    \begin{aligned}
        (\ldots) \ast S^{ij} \ast (\ldots) \ast S^{mn} \ast (\ldots),
    \end{aligned}
\end{equation}
where the ellipsis $(...)$ represents irrelevant part to divergences. For convenience, we define the following propagators
\begin{equation}
  \begin{aligned}
      S_2^{i j}& (x) 
      :=  X^{\mu\nu}(x) gG_{\mu \nu}^a T^a_{i j} 
       = \frac{\Gamma \left( \frac{d}{2} - 1 \right) \gamma^{\mu}
     \slashed{x} \gamma^{\nu}}{16 \pi^{d / 2} (- x^2)^{d / 2 - 1}} g G_{\mu \nu}^a
     T^a_{i j}, 
     \end{aligned}
\end{equation}
\begin{equation}
  \begin{aligned}
     S_{4b}^{i j}& (x)
     := Z^{\mu\nu\rho\sigma}(x) g G_{\mu \nu ; \rho \sigma}^a T^a_{i j} \\
      = & \frac{- \Gamma \left( \frac{d}{2} - 2 \right) \left(
     {2 g^{\{ \mu \rho} x^{ \sigma \}}}
      + g^{\{ \mu \rho} \gamma^{ \sigma \}} \slashed{x} \right)
      \gamma^{\nu}}{2^8 \times 3 \pi^{d / 2} (- x^2)^{d / 2 - 2}} g G_{\mu \nu ;
      \rho \sigma}^a T^a_{i j} ,
     \end{aligned}
\end{equation}
\begin{equation}
  \begin{aligned}
      S^{i j}_{4a}& (x) 
       :=   g^2 G^a_{\mu\nu}G^b_{\rho\sigma}(T^aT^b)_{ij} Y^{\mu\nu\rho\sigma}(x)\\
      = &  g^2 G^a_{\mu \nu} G^b_{\rho \sigma} (T^a T^b)_{i j}
   \Bigg[ \frac{- \text{i} \Gamma \left( \frac{d}{2} - 2 \right) g^{\nu \sigma}
      \gamma^{\mu} \slashed{x} \gamma^{\rho}}{192 \pi^{d / 2} (- x^2)^{d / 2 - 2}}\\
      + & \frac{\text{i} \Gamma \left( \frac{d}{2} - 2 \right) \left( - 6
     g^{\nu \sigma} \slashed{x} \gamma^{\mu \rho} - 4 x^{\sigma} \gamma^{\mu \nu
     \rho}  \right)}{2^8 \times 3 \pi^{d / 2} (- x^2)^{d / 2 - 2}}   \\
     +   & \frac{\text{i} \Gamma \left( \frac{d}{2} - 2 \right) \left(  6 x^{\mu} \gamma^{\nu \rho \sigma} - 4 x^{\nu} \gamma^{\mu \sigma
     \rho} {+ 3 \gamma^{\mu} \slashed{x} \gamma^{\nu \rho
     \sigma}} \right)}{2^8 \times 3 \pi^{d / 2} (- x^2)^{d / 2 - 2}} \Bigg] ,
    \end{aligned}
\end{equation}
in which the subscript 2/4 represents the dimension of $S^{ij}(x)$. It is very clear that $S^{ij}_{4a}(x)$ and $S^{ij}_{4b}(x)$ are divergent with the factor $\Gamma(d/2-2)$ when $d=4$, which will appear in the Feynman diagrams Fig.~\ref{fig:Feynman4d} and Fig.~\ref{fig:Feynman4f} respectively. The sum of these two diagrams is shown in Fig.~\ref{fig:IR1} with the expression 
\begin{widetext}
\begin{equation}
    \begin{aligned} 
    (\ldots) &\ast S^{i j}_{4a}(x) \ast (\ldots) S^{m n}_2 \ast (\ldots) + (\ldots)
   \ast S^{i j}_{4b}(x) \ast (\ldots) S^{m n}_2 \ast (\ldots) =     \label{A.3}  
    \end{aligned}
\end{equation}
\begin{equation*}
    (\ldots)\ast [(T^b T^c)_{i j} g^2 G^b_{\mu_1 \nu_1} G^c_{\rho_1 \sigma_1} Y^{\mu_1 \nu_1 \rho_1 \sigma_1}(x) +T^a_{i j} gG^a_{\mu_2 \nu_2 ; \rho_2 \sigma_2} Z^{\mu_2 \nu_2 \rho_2 \sigma_2}(x) ] \ast (\ldots) \ast T^d_{m n} gG^d_{\mu \nu} X^{\mu\nu} \ast (\ldots).
\end{equation*}
\end{widetext}
We define following structures for the dimension-6 condensates
\begin{equation}
    \begin{aligned} 
   \langle G^a_{\mu \nu} G^b_{\alpha \beta} G^c_{\rho \sigma} \rangle & :=   f^{abc} W_{\mu\nu;\alpha\beta;\rho\sigma},\\
  \langle G^a_{\mu \nu ; \rho \sigma} G^b_{\alpha
  \beta} \rangle & := \delta^{ab} U_{\mu\nu\alpha\beta;\rho\sigma} ,
    \end{aligned}
\end{equation}
where 
\begin{equation}
    \begin{aligned}
  &g^3  W_{\mu\nu;\alpha\beta;\rho\sigma}  = 
   \frac{-\langle g^3 f G^3\rangle}{24(D-2)(D-1)D} \\&[g_{\beta\nu}(g_{\alpha\rho}g_{\mu\sigma}-g_{\alpha\sigma}g_{\mu\rho})+ g_{\alpha\nu}(g_{\beta\sigma}g_{\mu\rho}-g_{\beta\rho}g_{\mu\sigma}) \\
  & + g_{\beta\mu}(g_{\alpha\sigma}g_{\nu\rho}-g_{\alpha\rho}g_{\nu\sigma}) +g_{\alpha\mu}(g_{\beta\rho}g_{\nu\sigma}-g_{\beta\sigma}g_{\nu\rho})].
    \end{aligned}
\end{equation}
The specific expression of $U_{\mu\nu\alpha\beta;\rho\sigma}$ is given by Eqs.~\eqref{eq:njl} and ~\eqref{eq:njl11}. Then we have
\begin{equation}
    \begin{aligned} 
   &(T^b T^c)_{i j} T^d_{m n} \langle G^b_{\mu_1 \nu_1} G^c_{\rho_1 \sigma_1}
  G^d_{\mu \nu} \rangle \\ =&  (T^b T^c)_{i j} T^d_{m n} f^{b c d} W_{\mu_1 \nu_1;\rho_1\sigma_1;\mu\nu},\label{A.6}
    \end{aligned}
\end{equation}
\begin{equation}
    \begin{aligned} 
   &T^a_{i j} T^d_{m n} \langle G^a_{\mu_2 \nu_2 ; \rho_2 \sigma_2} G^d_{\mu
  \nu} \rangle    =T^a_{i j} T^a_{m n} U_{\mu_2\nu_2\rho_2\sigma_2;\mu\nu}, \label{A.7}
    \end{aligned}
\end{equation}
and 
\begin{equation}
    \begin{aligned} 
 (T^b T^c)_{i j} T^d_{m n} f^{b c d}  =  \frac{3 \text{i}}{2}
   T^d_{m n} T^d_{i j}, \label{A.8}
   \end{aligned}
\end{equation}
where $f^{bce}$ and $d^{bce}$ are antisymmetric structure and symmetric constants of SU(3) group, respectively. Substituting Eqs.~\eqref{A.6}-\eqref{A.8} into Eq.~\eqref{A.3}
\begin{equation}
    \begin{aligned} 
    (...)\ast T^d_{m n}& T^d_{i j}  [ \frac{3 \text{i}}{2}  W_{\mu_1 \nu_1;\rho_1\sigma_1;\mu\nu}  g^3 Y^{\mu_1 \nu_1 \rho_1\sigma_1}(x) \\ +& U_{\mu_2\nu_2\rho_2\sigma_2;\mu\nu}  g^2 Z^{\mu_2   \nu_2\rho_2\sigma_2}(x) ] \ast (\dots) \ast X^{\mu\nu},
    \end{aligned}
\end{equation}
in which the result in the square brackets can be calculated as 
\begin{equation}
    \begin{aligned} 
 &\frac{\langle g^3 f G^3 \rangle \Gamma \left( \frac{d}{2} - 2 \right) (d -
   4) \left( \gamma_{\mu} \slashed{x} \gamma_{\nu} - \gamma_{\nu} \slashed{x}
   \gamma_{\mu} \right)}{3072 (d - 2) (d - 1) d \pi^{d / 2} (- x^2)^{d / 2 -
   2}} \\  =  & \frac{\langle g^3 f G^3 \rangle \Gamma \left( \frac{d}{2} - 1 \right)
   \left( \gamma_{\mu} \slashed{x} \gamma_{\nu} - \gamma_{\nu} \slashed{x}
   \gamma_{\mu} \right)}{1536 (d - 2) (d - 1) d \pi^{d / 2} (- x^2)^{d / 2 -
   2}} . 
   \end{aligned}
\end{equation}
It is fantastic that this result is finite, showing that the divergences in Fig.~\ref{fig:Feynman4d} and Fig.~\ref{fig:Feynman4f} exactly cancel out for each other. 

To ensure such a cancellation for the calculation of Fig.~\ref{fig:Feynman4e}, it is 
important to use the $d$-dimensional propagator and $D$-dimensional condensates, but not any one in the 4-dimensional. The propagator containing $G^a_{\mu\nu;\rho}$ can be written as
\begin{equation}
    \begin{aligned} 
    S_3^{ij} (x)& = S_{3a}^{ij} (x) + S_{3b}^{ij} (x) := J^{\mu\nu\rho}_0  gG_{\mu \nu ; \rho}^a T^a_{i j},
   \end{aligned}
\end{equation}
\begin{equation}
    \begin{aligned} 
 S_{3a}^{ij} (x)& := J^{\mu\nu\rho}_1  gG_{\mu \nu ; \rho}^a T^a_{i j}  \\ &= \frac{\Gamma \left( \frac{d}{2} - 1 \right)g G_{\mu
   \nu ; \rho}^a T^a_{i j}}{48
   \pi^{d / 2} (- x^2)^{d / 2 - 1}} \left( x_{\mu} \gamma_{\rho} \slashed{x}
   \gamma_{\nu} + x_{\rho} \gamma_{\mu} \slashed{x} \gamma_{\nu} \right)  , 
   \end{aligned}
\end{equation}
\begin{equation}
    \begin{aligned} 
 S_{3b}^{ij}(x)  & := J^{\mu\nu\rho}_2  gG_{\mu \nu ; \rho}^a T^a_{i j} (x)  \\ & = \frac{\Gamma \left( \frac{d}{2} - 2 \right)g G_{\mu \nu ;
   \rho}^a T^a_{i j}}{96
   \pi^{d / 2} (- x^2)^{d / 2 - 2}} {(\gamma_{\mu \rho
   \nu} + \gamma_{\rho \mu \nu} - 4 g_{\mu \rho} \gamma_{\nu})} ,
   \end{aligned}
\end{equation}
where $J^{\mu\nu\rho}_0, J^{\mu\nu\rho}_1, J^{\mu\nu\rho}_2$ are corresponding tensor structures and $J^{\mu\nu\rho}_0=J^{\mu\nu\rho}_1+J^{\mu\nu\rho}_2$. 
It is clear that only $S_{3b}^{ij}$ is divergent while $S_{3a}^{ij}$ remains finite. The most divergent part of Fig.~\ref{fig:Feynman4e} can be written as
\begin{widetext}
    \begin{equation}
  \begin{aligned}  
     (\ldots) \ast S^{i j}_{3b} \ast (\ldots) S^{m n}_{3b} \ast (\ldots) & = (\ldots)
   \ast J_2^{\mu_1 \nu_1 \rho_1} T^a_{ij} g G_{\mu_1 \nu_1 ; \rho_1}^a \ast
   (\ldots) \ast J_2^{\mu \nu \rho} T^b_{m n} gG_{\mu \nu ; \rho}^b \ast (\ldots)\\ 
  & = (\ldots) \ast J_2^{\mu_1 \nu_1 \rho_1} T^a_{i j} T^b_{m n} g^2\langle G_{\mu_1 \nu_1 ; \rho_1}^a G_{\mu \nu ; \rho}^b \rangle \ast (\ldots) \ast J_2^{\mu \nu \rho} \ast (\ldots) ,
   \end{aligned}
\end{equation}
\end{widetext}
in which both $J_2^{\mu_1 \nu_1 \rho_1}$ and $J_2^{\mu \nu \rho}$ contain the divergent factor $\Gamma(\frac{d}{2}-2)$. However, such a divergence is trivial since the other part in the calculation is zero
\begin{equation}
    \begin{aligned} 
 &J_2^{\mu_1 \nu_1 \rho_1} T^a_{i j} T^b_{m n} g^2 \langle G_{\mu_1 \nu_1 ;
   \rho_1}^a G_{\mu \nu ; \rho}^b \rangle \\ =& T^a_{i j} T^b_{m n} \frac{\langle g^3 f G^3 \rangle
   \Gamma \left( \frac{d}{2} - 2 \right) (D - d) (g_{\rho \mu} \gamma_{\nu} -
   g_{\rho \nu} \gamma_{\mu})}{128 (D - 1) D (D^2 - 4) \pi^{d / 2} (- x^2)^{d
   / 2 - 2}}  \\ =& 0, 
   \end{aligned}
\end{equation}
where we use $D=d$ in the last step.
\section{Two-point correlation functions for all interpolating currents}\label{appendix: ope}
In this appendix, we present the correlation functions after Borel transformation up to dimension-8 for the interpolating currents in Eq.~\eqref{eq:currents}. 
\begin{widetext}
    \begin{eqnarray*}
  \Pi_{T_6} & = & \frac{3 M_B^{10}}{160 \pi^6} + \left( \frac{11 \langle g^2
  G^2 \rangle}{384 \pi^6}  + \frac{m \langle \overline{s} s \rangle}{4 \pi^4}
  \right) M_B^6 + \left( \frac{7 m \langle g \overline{s} \sigma G s
  \rangle}{32 \pi^4} - \frac{\langle g^3 f G^3 \rangle}{16 \pi^6} + \frac{7
  g^2 \langle \overline{q} q \rangle^2}{36 \pi^4} + \frac{7 g^2 \langle
  \overline{s} s \rangle^2}{108 \pi^4} \right) M_B^4\\
  & + & \frac{11 m \langle g^2 G^2 \rangle \langle \overline{s} s
  \rangle}{192 \pi^4} M_B^2.  %+ \frac{16 m g^2 \langle \overline{s} s \rangle
  %\langle \overline{q} q \rangle^2}{81 \pi^2} + \frac{16}{3} m \langle
  %\overline{q} q \rangle^3 - \frac{11 m \langle g^2 G^2 \rangle \langle g
  %\overline{s} \sigma G s \rangle}{1152 \pi^4},
\end{eqnarray*}
\begin{eqnarray*}
  \Pi_{T_3} & = & \frac{3 M_B^{10}}{320 \pi^6} + \left( \frac{\langle g^2 G^2
  \rangle}{384 \pi^6}  + \frac{m \langle \overline{s} s \rangle}{8 \pi^4}
  \right) M_B^6 - \left( \frac{\langle g^3 f G^3 \rangle}{64 \pi^6} + \frac{m
  \langle g \overline{s} \sigma G s \rangle}{32 \pi^4} + \frac{g^2 \langle
  \overline{q} q \rangle^2}{36 \pi^4} + \frac{g^2 \langle \overline{s} s
  \rangle^2}{108 \pi^4} \right) M_B^4\\
  & + & \frac{m \langle g^2 G^2 \rangle \langle \overline{s} s \rangle}{192
  \pi^4} M_B^2. %- \frac{m g^2 \langle \overline{s} s \rangle \langle
  %\overline{q} q \rangle^2}{81 \pi^2} - \frac{m \langle g^2 G^2 \rangle
  %\langle g \overline{s} \sigma G s \rangle}{1152 \pi^4} + \frac{8}{3} m
  %\langle \overline{q} q \rangle^3,
\end{eqnarray*}
\begin{eqnarray*}
  \Pi_{S_6} & = & \frac{M_B^{10}}{1280 \pi^6} + \left( \frac{m \langle
  \overline{s} s \rangle}{96 \pi^4} - \frac{m \langle \overline{q} q
  \rangle}{48 \pi^4} - \frac{\langle g^2 G^2 \rangle}{3072 \pi^6} \right)
  {M^6_B} \\
  & + & \left( \frac{m \langle g \overline{q} G \sigma q \rangle}{64 \pi^4} -
  \frac{5 \langle g^3 f G^3 \rangle}{13824 \pi^6} - \frac{\langle \overline{q}
  q \rangle^2}{12 \pi^2} + \frac{\langle \overline{q} q \rangle \langle
  \overline{s} s \rangle}{12 \pi^2} - \frac{7 m \langle g \overline{s} \sigma
  G s \rangle}{768 \pi^4} - \frac{7 g^2 \langle \overline{q} q \rangle^2}{864
  \pi^4} - \frac{7 g^2 \langle \overline{s} s \rangle^2}{2592 \pi^4} \right)
  M_B^4\\
  & + & \left( \frac{\langle \overline{q} q \rangle \langle g \overline{q}
  \sigma G q \rangle}{12 \pi^2} - \frac{m \langle g^2 G^2 \rangle \langle
  \overline{q} q \rangle}{2304 \pi^4} - \frac{\langle g \overline{s} \sigma G
  s \rangle \langle \overline{q} q \rangle}{24 \pi^2} - \frac{m \langle g^2
  G^2 \rangle \langle \overline{s} s \rangle}{1536 \pi^4} - \frac{\langle g
  \overline{q} \sigma G q \rangle \langle \overline{s} s \rangle}{24 \pi^2}
  \right) M_B^2.
  %& + & \frac{2}{9} m \langle \overline{q} q \rangle^3 - \frac{1}{9} m
  %\langle \overline{q} q \rangle^2 \langle \overline{s} s \rangle - \frac{m
  %\langle g^2 G^2 \rangle \langle g \overline{q} \sigma G q \rangle}{3072
  %\pi^4} + \frac{m \langle g^2 G^2 \rangle \langle g \overline{s} \sigma G s
  %\rangle}{9216 \pi^4} - \frac{139 \langle g \overline{q} \sigma G q
  %\rangle^2}{9216 \pi^2}\\
  %& - & \frac{5 \langle g^2 G^2 \rangle \langle \overline{q} q
  %\rangle^2}{1728 \pi^2} + \frac{139 \langle g \overline{q} \sigma G q \rangle
  %\langle g \overline{s} \sigma G s \rangle}{9216 \pi^2} - \frac{g^2 m \langle
  %\overline{q} q \rangle^3}{972 \pi^2} + \frac{5 \langle g^2 G^2 \rangle
  %\langle \overline{q} q \rangle \langle \overline{s} s \rangle}{1728 \pi^2} -
  %\frac{g^2 m \langle \overline{q} q \rangle^2 \langle \overline{s} s
  %\rangle}{162 \pi^2}\\
  %& + & \frac{g^2 m \langle \overline{q} q \rangle \langle \overline{s} s
  %\rangle^2}{972 \pi^2},
\end{eqnarray*}
\begin{eqnarray*}
  \Pi_{S_3} & = & \frac{M_B^{10}}{2560 \pi^6} + \left( \frac{m \langle
  \overline{s} s \rangle}{192 \pi^4} - \frac{m \langle \overline{q} q
  \rangle}{96 \pi^4} + \frac{\langle g^2 G^2 \rangle}{3072 \pi^6} \right)
  {M^6_B} \\
  & + & \left( \frac{\langle g^3 f G^3 \rangle}{6912 \pi^6} + \frac{m \langle
  g \overline{q} G \sigma q \rangle}{128 \pi^4} + \frac{m \langle g
  \overline{s} \sigma G s \rangle}{768 \pi^4} - \frac{\langle \overline{q} q
  \rangle^2}{24 \pi^2} + \frac{\langle \overline{q} q \rangle \langle
  \overline{s} s \rangle}{24 \pi^2} + \frac{g^2 \langle \overline{q} q
  \rangle^2}{864 \pi^4} + \frac{g^2 \langle \overline{s} s \rangle^2}{2592
  \pi^4} \right) M_B^4\\
  & + & \left( - \frac{5 m \langle g^2 G^2 \rangle \langle \overline{q} q
  \rangle}{2304 \pi^4} + \frac{\langle \overline{q} q \rangle \langle g
  \overline{q} \sigma G q \rangle}{24 \pi^2} - \frac{\langle g \overline{q}
  \sigma G q \rangle \langle \overline{s} s \rangle}{48 \pi^2} - \frac{\langle
  g \overline{s} \sigma G s \rangle \langle \overline{q} q \rangle}{48 \pi^2}
  + \frac{m \langle g^2 G^2 \rangle \langle \overline{s} s \rangle}{1536
  \pi^4} \right) M_B^2.
  %& + & \frac{2}{9} m \langle \overline{q} q \rangle^3 + \frac{m \langle g^2
  %G^2 \rangle \langle g \overline{q} \sigma G q \rangle}{3072 \pi^4} - \frac{m
  %\langle g^2 G^2 \rangle \langle g \overline{s} \sigma G s \rangle}{9216
  %\pi^4} + \frac{59 \langle g \overline{q} \sigma G q \rangle \langle g
  %\overline{s} \sigma G s \rangle}{9216 \pi^2} - \frac{59 \langle g
  %\overline{q} \sigma G q \rangle^2}{9216 \pi^2}\\
  %& - & \frac{7 \langle g^2 G^2 \rangle \langle \overline{q} q
  %\rangle^2}{1728 \pi^2} - \frac{5 g^2 m \langle \overline{q} q \rangle^3}{972
  %\pi^2} + \frac{7 \langle g^2 G^2 \rangle \langle \overline{q} q \rangle
  %\langle \overline{s} s \rangle}{1728 \pi^2} - \frac{1}{18} m \langle
  %\overline{q} q \rangle^2 \langle \overline{s} s \rangle + \frac{g^2 m
  %\langle \overline{q} q \rangle^2 \langle \overline{s} s \rangle}{648 \pi^2}
  %+ \frac{g^2 m \langle \overline{q} q \rangle \langle \overline{s} s
  %\rangle^2}{1944 \pi^2},
\end{eqnarray*}
\begin{eqnarray*}
  \Pi_{V_6} & = & \frac{M_B^{10}}{1280 \pi^6} + \left( \frac{m \langle
  \overline{s} s \rangle}{96 \pi^4} + \frac{m \langle \overline{q} q
  \rangle}{48 \pi^4} - \frac{\langle g^2 G^2 \rangle}{3072 \pi^6} \right)
  {M^6_B} \\
  & + & \left( \frac{- 5 \langle g^3 f G^3 \rangle}{13824 \pi^6} - \frac{m
  \langle g \overline{q} G \sigma q \rangle}{64 \pi^4} - \frac{7 m \langle g
  \overline{s} \sigma G s \rangle}{768 \pi^4} + \frac{\langle \overline{q} q
  \rangle^2}{12 \pi^2} - \frac{\langle \overline{q} q \rangle \langle
  \overline{s} s \rangle}{12 \pi^2} - \frac{7 g^2 \langle \overline{q} q
  \rangle^2}{864 \pi^4} - \frac{7 g^2 \langle \overline{s} s \rangle^2}{2592
  \pi^4} \right) M_B^4\\
  & + & \left( \frac{m \langle g^2 G^2 \rangle \langle \overline{q} q
  \rangle}{2304 \pi^4} - \frac{\langle \overline{q} q \rangle \langle g
  \overline{q} \sigma G q \rangle}{12 \pi^2} + \frac{\langle g \overline{s}
  \sigma G s \rangle \langle \overline{q} q \rangle}{24 \pi^2} + \frac{\langle
  g \overline{q} \sigma G q \rangle \langle \overline{s} s \rangle}{24 \pi^2}
  - \frac{m \langle g^2 G^2 \rangle \langle \overline{s} s \rangle}{1536
  \pi^4} \right) M_B^2.
 % & + & \frac{2}{9} m \langle \overline{q} q \rangle^3 + \frac{m \langle g^2
  %G^2 \rangle \langle g \overline{q} \sigma G q \rangle}{3072 \pi^4} + \frac{m
 % \langle g^2 G^2 \rangle \langle g \overline{s} \sigma G s \rangle}{9216
  %\pi^4} - \frac{139 \langle g \overline{q} \sigma G q \rangle \langle g
  %\overline{s} \sigma G s \rangle}{9216 \pi^2} + \frac{139 \langle g
 % \overline{q} \sigma G q \rangle^2}{9216 \pi^2}\\
 % & + & \frac{5 \langle g^2 G^2 \rangle \langle \overline{q} q
  %\rangle^2}{1728 \pi^2} + \frac{g^2 m \langle \overline{q} q \rangle^3}{972
  %%\pi^2} - \frac{5 \langle g^2 G^2 \rangle \langle \overline{q} q \rangle
  %\langle \overline{s} s \rangle}{1728 \pi^2} + \frac{1}{9} m \langle
  %\overline{q} q \rangle^2 \langle \overline{s} s \rangle - \frac{g^2 m
  %\langle \overline{q} q \rangle^2 \langle \overline{s} s \rangle}{162 \pi^2}
  %- \frac{g^2 m \langle \overline{q} q \rangle \langle \overline{s} s
  %\rangle^2}{972 \pi^2},
\end{eqnarray*}
\begin{eqnarray*}
  \Pi_{V_3} & = & \frac{M_B^{10}}{2560 \pi^6} + \left( \frac{m \langle
  \overline{s} s \rangle}{192 \pi^4} + \frac{m \langle \overline{q} q
  \rangle}{96 \pi^4} + \frac{\langle g^2 G^2 \rangle}{3072 \pi^6} \right)
  {M^6_B} \\
  & + & \left( \frac{\langle g^3 f G^3 \rangle}{6912 \pi^6} - \frac{m \langle
  g \overline{q} G \sigma q \rangle}{128 \pi^4} + \frac{m \langle g
  \overline{s} \sigma G s \rangle}{768 \pi^4} + \frac{\langle \overline{q} q
  \rangle^2}{24 \pi^2} - \frac{\langle \overline{q} q \rangle \langle
  \overline{s} s \rangle}{24 \pi^2} + \frac{g^2 \langle \overline{q} q
  \rangle^2}{864 \pi^4} + \frac{g^2 \langle \overline{s} s \rangle^2}{2592
  \pi^4} \right) M_B^4\\
  & + & \left( \frac{5 m \langle g^2 G^2 \rangle \langle \overline{q} q
  \rangle}{2304 \pi^4} - \frac{\langle \overline{q} q \rangle \langle g
  \overline{q} \sigma G q \rangle}{24 \pi^2} + \frac{\langle g \overline{s}
  \sigma G s \rangle \langle \overline{q} q \rangle}{48 \pi^2} + \frac{\langle
  g \overline{q} \sigma G q \rangle \langle \overline{s} s \rangle}{48 \pi^2}
  + \frac{m \langle g^2 G^2 \rangle \langle \overline{s} s \rangle}{1536
  \pi^4} \right) M_B^2.
 % & + & \frac{1}{9} m \langle \overline{q} q \rangle^3 - \frac{m \langle g^2
 % G^2 \rangle \langle g \overline{q} \sigma G q \rangle}{3072 \pi^4} - \frac{m
 % \langle g^2 G^2 \rangle \langle g \overline{s} \sigma G s \rangle}{9216
 % \pi^4} - \frac{59 \langle g \overline{q} \sigma G q \rangle \langle g
 % \overline{s} \sigma G s \rangle}{9216 \pi^2} + \frac{59 \langle g
 % \overline{q} \sigma G q \rangle^2}{9216 \pi^2}\\
%  & + & \frac{7 \langle g^2 G^2 \rangle \langle \overline{q} q
%  \rangle^2}{1728 \pi^2} + \frac{5 g^2 m \langle \overline{q} q \rangle^3}{972
 % \pi^2} - \frac{7 \langle g^2 G^2 \rangle \langle \overline{q} q \rangle
%  \langle \overline{s} s \rangle}{1728 \pi^2} + \frac{m \langle \overline{q} q
%  \rangle^2 \langle \overline{s} s \rangle}{18} + \frac{g^2 m \langle
 % \overline{q} q \rangle^2 \langle \overline{s} s \rangle}{648 \pi^2} -
%  \frac{g^2 m \langle \overline{q} q \rangle \langle \overline{s} s
%  \rangle^2}{1944 \pi^2},
\end{eqnarray*}
\begin{eqnarray*}
  \Pi_{P_6} & = & \frac{M_B^{10}}{320 \pi^6} + \left( \frac{m \langle
  \overline{s} s \rangle}{24 \pi^4} - \frac{m \langle \overline{q} q
  \rangle}{24 \pi^4} + \frac{5 \langle g^2 G^2 \rangle}{1536 \pi^6} \right)
  {M^6_B} \\
  & + & \left( - \frac{113 \langle g^3 f G^3 \rangle}{13824 \pi^6} - \frac{m
  \langle g \overline{q} G \sigma q \rangle}{128 \pi^4} + \frac{7 m \langle g
  \overline{s} \sigma G s \rangle}{384 \pi^4} - \frac{\langle \overline{q} q
  \rangle^2}{6 \pi^2} + \frac{\langle \overline{q} q \rangle \langle
  \overline{s} s \rangle}{6 \pi^2} + \frac{7 g^2 \langle \overline{q} q
  \rangle^2}{432 \pi^4} + \frac{7 g^2 \langle \overline{s} s \rangle^2}{1296
  \pi^4} \right) M_B^4\\
  & + & \left( - \frac{7 m \langle g^2 G^2 \rangle \langle \overline{q} q
  \rangle}{1152 \pi^4} - \frac{\langle \overline{q} q \rangle \langle g
  \overline{q} \sigma G q \rangle}{24 \pi^2} + \frac{\langle g \overline{s}
  \sigma G s \rangle \langle \overline{q} q \rangle}{48 \pi^2} + \frac{\langle
  g \overline{q} \sigma G q \rangle \langle \overline{s} s \rangle}{48 \pi^2}
  + \frac{5 m \langle g^2 G^2 \rangle \langle \overline{s} s \rangle}{768
  \pi^4} \right) M_B^2.
%  & + & \frac{8}{9} m \langle \overline{q} q \rangle^3 + \frac{m \langle g^2
%  G^2 \rangle \langle g \overline{q} \sigma G q \rangle}{1536 \pi^4} - \frac{5
%  m \langle g^2 G^2 \rangle \langle g \overline{s} \sigma G s \rangle}{4608
%  \pi^4} - \frac{5 \langle g \overline{q} \sigma G q \rangle \langle g
%  \overline{s} \sigma G s \rangle}{1536 \pi^2} - \frac{11 \langle g^2 G^2
%  \rangle \langle \overline{q} q \rangle^2}{864 \pi^2}\\
%  & + & \frac{5 \langle g \overline{q} \sigma G q \rangle^2}{1536 \pi^2} -
%  \frac{19 g^2 m \langle \overline{q} q \rangle^3}{486 \pi^2} + \frac{11
%  \langle g^2 G^2 \rangle \langle \overline{q} q \rangle \langle \overline{s}
%  s \rangle}{864 \pi^2} - \frac{2}{9} m \langle \overline{q} q \rangle^2
%  \langle \overline{s} s \rangle + \frac{g^2 m \langle \overline{q} q
%  \rangle^2 \langle \overline{s} s \rangle}{54 \pi^2} + \frac{g^2 m \langle
%  \overline{q} q \rangle \langle \overline{s} s \rangle^2}{486 \pi^2},
\end{eqnarray*}
\begin{eqnarray*}
  \Pi_{P_3} & = & \frac{M_B^{10}}{640 \pi^6} + \left( \frac{m \langle
  \overline{s} s \rangle}{48 \pi^4} - \frac{m \langle \overline{q} q
  \rangle}{48 \pi^4} + \frac{5 \langle g^2 G^2 \rangle}{1536 \pi^6} \right)
  {M^6_B} \\
  & + & \left( - \frac{113 \langle g^3 f G^3 \rangle}{13824 \pi^6} - \frac{3
  m \langle g \overline{q} G \sigma q \rangle}{128 \pi^4} + \frac{11 m \langle
  g \overline{s} \sigma G s \rangle}{384 \pi^4} - \frac{\langle \overline{q} q
  \rangle^2}{12 \pi^2} + \frac{\langle \overline{q} q \rangle \langle
  \overline{s} s \rangle}{12 \pi^2} + \frac{5 g^2 \langle \overline{q} q
  \rangle^2}{432 \pi^4} + \frac{5 g^2 \langle \overline{s} s \rangle^2}{1296
  \pi^4} \right) M_B^4\\
  & + & \left( - \frac{5 m \langle g^2 G^2 \rangle \langle \overline{q} q
  \rangle}{1152 \pi^4} - \frac{\langle \overline{q} q \rangle \langle g
  \overline{q} \sigma G q \rangle}{8 \pi^2} + \frac{\langle g \overline{s}
  \sigma G s \rangle \langle \overline{q} q \rangle}{16 \pi^2} + \frac{\langle
  g \overline{q} \sigma G q \rangle \langle \overline{s} s \rangle}{16 \pi^2}
  + \frac{5 m \langle g^2 G^2 \rangle \langle \overline{s} s \rangle}{768
  \pi^4} \right) M_B^2.
%  & + & \frac{4}{9} m \langle \overline{q} q \rangle^3 + \frac{m \langle g^2
%  G^2 \rangle \langle g \overline{q} \sigma G q \rangle}{1536 \pi^4} - \frac{5
%  m \langle g^2 G^2 \rangle \langle g \overline{s} \sigma G s \rangle}{4608
%  \pi^4} - \frac{7 \langle g \overline{q} \sigma G q \rangle \langle g
%  \overline{s} \sigma G s \rangle}{512 \pi^2} - \frac{7 \langle g^2 G^2
%  \rangle \langle \overline{q} q \rangle^2}{864 \pi^2}\\
% & + & \frac{7 \langle g \overline{q} \sigma G q \rangle^2}{512 \pi^2} -
%  \frac{17 g^2 m \langle \overline{q} q \rangle^3}{486 \pi^2} + \frac{7
%  \langle g^2 G^2 \rangle \langle \overline{q} q \rangle \langle \overline{s}
 % s \rangle}{864 \pi^2} - \frac{1}{9} m \langle \overline{q} q \rangle^2
 % \langle \overline{s} s \rangle + \frac{g^2 m \langle \overline{q} q
 % \rangle^2 \langle \overline{s} s \rangle}{81 \pi^2} + \frac{g^2 m \langle
 % \overline{q} q \rangle \langle \overline{s} s \rangle^2}{972 \pi^2},
\end{eqnarray*}
\begin{eqnarray*}
  \Pi_{A_6} & = & \frac{M_B^{10}}{320 \pi^6} + \left( \frac{m \langle
  \overline{s} s \rangle}{24 \pi^4} + \frac{m \langle \overline{q} q
  \rangle}{24 \pi^4} + \frac{5 \langle g^2 G^2 \rangle}{1536 \pi^6} \right)
  {M^6_B} \\
  & + & \left( - \frac{113 \langle g^3 f G^3 \rangle}{13824 \pi^6} + \frac{m
  \langle g \overline{q} G \sigma q \rangle}{128 \pi^4} + \frac{7 m \langle g
  \overline{s} \sigma G s \rangle}{384 \pi^4} + \frac{\langle \overline{q} q
  \rangle^2}{6 \pi^2} - \frac{\langle \overline{q} q \rangle \langle
  \overline{s} s \rangle}{6 \pi^2} + \frac{7 g^2 \langle \overline{q} q
  \rangle^2}{432 \pi^4} + \frac{7 g^2 \langle \overline{s} s \rangle^2}{1296
  \pi^4} \right) M_B^4\\
  & + & \left( \frac{7 m \langle g^2 G^2 \rangle \langle \overline{q} q
  \rangle}{1152 \pi^4} + \frac{\langle \overline{q} q \rangle \langle g
  \overline{q} \sigma G q \rangle}{24 \pi^2} - \frac{\langle g \overline{s}
  \sigma G s \rangle \langle \overline{q} q \rangle}{48 \pi^2} - \frac{\langle
  g \overline{q} \sigma G q \rangle \langle \overline{s} s \rangle}{48 \pi^2}
  + \frac{5 m \langle g^2 G^2 \rangle \langle \overline{s} s \rangle}{768
  \pi^4} \right) M_B^2.
 % & + & \frac{8}{9} m \langle \overline{q} q \rangle^3 - \frac{m \langle g^2
%%  G^2 \rangle \langle g \overline{q} \sigma G q \rangle}{1536 \pi^4} - \frac{5
%  m \langle g^2 G^2 \rangle \langle g \overline{s} \sigma G s \rangle}{4608
 %% \pi^4} + \frac{5 \langle g \overline{q} \sigma G q \rangle \langle g
 % \overline{s} \sigma G s \rangle}{1536 \pi^2} + \frac{11 \langle g^2 G^2
 % \rangle \langle \overline{q} q \rangle^2}{864 \pi^2}\\
 % & - & \frac{5 \langle g \overline{q} \sigma G q \rangle^2}{1536 \pi^2} +
 % \frac{19 g^2 m \langle \overline{q} q \rangle^3}{486 \pi^2} - \frac{11
 % \langle g^2 G^2 \rangle \langle \overline{q} q \rangle \langle \overline{s}
 % s \rangle}{864 \pi^2} + \frac{2}{9} m \langle \overline{q} q \rangle^2
 % \langle \overline{s} s \rangle + \frac{g^2 m \langle \overline{q} q
 % \rangle^2 \langle \overline{s} s \rangle}{54 \pi^2} - \frac{g^2 m \langle
 % \overline{q} q \rangle \langle \overline{s} s \rangle^2}{486 \pi^2},
\end{eqnarray*}
\begin{eqnarray*}
  \Pi_{A_3} & = & \frac{M_B^{10}}{640 \pi^6} + \left( \frac{m \langle
  \overline{s} s \rangle}{48 \pi^4} + \frac{m \langle \overline{q} q
  \rangle}{48 \pi^4} + \frac{\langle g^2 G^2 \rangle}{1536 \pi^6} \right)
  {M^6_B} \\
  & + & \left( - \frac{25 \langle g^3 f G^3 \rangle}{13824 \pi^6} - \frac{m
  \langle g \overline{q} G \sigma q \rangle}{128 \pi^4} - \frac{m \langle g
  \overline{s} \sigma G s \rangle}{384 \pi^4} + \frac{\langle \overline{q} q
  \rangle^2}{12 \pi^2} - \frac{\langle \overline{q} q \rangle \langle
  \overline{s} s \rangle}{12 \pi^2} - \frac{g^2 \langle \overline{q} q
  \rangle^2}{432 \pi^4} - \frac{g^2 \langle \overline{s} s \rangle^2}{1296
  \pi^4} \right) M_B^4\\
  & + & \left( - \frac{m \langle g^2 G^2 \rangle \langle \overline{q} q
  \rangle}{1152 \pi^4} - \frac{\langle \overline{q} q \rangle \langle g
  \overline{q} \sigma G q \rangle}{24 \pi^2} + \frac{\langle g \overline{s}
  \sigma G s \rangle \langle \overline{q} q \rangle}{48 \pi^2} + \frac{\langle
  g \overline{q} \sigma G q \rangle \langle \overline{s} s \rangle}{48 \pi^2}
  + \frac{m \langle g^2 G^2 \rangle \langle \overline{s} s \rangle}{768 \pi^4}
  \right) M_B^2.
 % & + & \frac{4}{9} m \langle \overline{q} q \rangle^3 + \frac{m \langle g^2
 % G^2 \rangle \langle g \overline{q} \sigma G q \rangle}{1536 \pi^4} - \frac{m
 % \langle g^2 G^2 \rangle \langle g \overline{s} \sigma G s \rangle}{4608
 % \pi^4} - \frac{11 \langle g \overline{q} \sigma G q \rangle \langle g
 % \overline{s} \sigma G s \rangle}{1536 \pi^2} + \frac{\langle g^2 G^2 \rangle
 % \langle \overline{q} q \rangle^2}{864 \pi^2}\\
 % & + & \frac{11 \langle g \overline{q} \sigma G q \rangle^2}{1536 \pi^2} -
%  \frac{13 g^2 m \langle \overline{q} q \rangle^3}{486 \pi^2} - \frac{\langle
 %% g^2 G^2 \rangle \langle \overline{q} q \rangle \langle \overline{s} s
 % \rangle}{864 \pi^2} + \frac{1}{9} m \langle \overline{q} q \rangle^2 \langle
%  \overline{s} s \rangle - \frac{g^2 m \langle \overline{q} q \rangle \langle
 % \overline{s} s \rangle^2}{972 \pi^2}.
\end{eqnarray*}
\end{widetext}

%\bibliographystyle{elsarticle-num}
%\bibliography{main}

\end{document}